\documentclass{article}

\usepackage{arxiv}

\usepackage{amsmath}		   % added for \begin{split}  -author

\usepackage[utf8]{inputenc} % allow utf-8 input
\usepackage[T1]{fontenc}    % use 8-bit T1 fonts
\usepackage{hyperref}       % hyperlinks
\usepackage{url}            % simple URL typesetting
\usepackage{booktabs}       % professional-quality tables
\usepackage{amsfonts}       % blackboard math symbols
\usepackage{nicefrac}       % compact symbols for 1/2, etc.
\usepackage{microtype}      % microtypography
\usepackage{cleveref}       % smart cross-referencing
\usepackage{lipsum}         % Can be removed after putting your text content
\usepackage{graphicx}
\usepackage{doi}

\usepackage[numbers, authoryear]{natbib} % Enables author-year citation
\renewcommand{\cite}{\citep} % Ensures citations appear as [Author Year]
\usepackage[ruled]{algorithm2e} 		% for algorithms
\usepackage{tabularx}				% for table formatting
\usepackage{makecell, multirow}		% for cell formatting
\usepackage{amsthm}					% proofs in appendices 

\usepackage{xcolor}

\title{Projective Displacement Mapping for Ray Traced Editable Surfaces}

\date{Feb 2, 2025}

\newif\ifuniqueAffiliation

\uniqueAffiliationtrue

\ifuniqueAffiliation 
\author{ \href{https://orcid.org/0000-0002-0449-981X}{\includegraphics[scale=0.06]{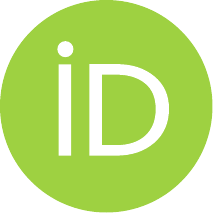}\hspace{1mm}Rama Carl Hoetzlein} \\
	Quanta Sciences\\
	Ithaca, NY 14850 \\
	\texttt{ramahoetzlein@gmail.com} \\
}
	
\else

\fi

\renewcommand{\shorttitle}{Projective Displacement Mapping for Ray Traced Editable Surfaces}

\hypersetup{
pdftitle={Projective Displacement Mapping for Ray Traced Editable Surfaces},
pdfsubject={q-bio.QM, q-bio.NC},
pdfauthor={Rama Carl Hoetzlein},
pdfkeywords={displacement mapping, 3D modeling, ray tracing, geometric detail},
}

\begin{document}
\maketitle

\begin{abstract}
Displacement mapping is an important tool for modeling detailed geometric features. We explore the problem of authoring complex surfaces while ray tracing interactively. Current techniques for ray tracing displaced surfaces rely on acceleration structures that require dynamic rebuilding when edited. These techniques are typically used for massive static scenes or the compression of detailed source assets. Our interest lies in modeling and look development of artistic features with real-time ray tracing. We demonstrate projective displacement mapping, a direct sampling method without a bottom-level acceleration structure combined with a top-level hardware BVH. Quality and performance are improved over existing methods with smoothed displaced normals, thin feature sampling, tight prism bounds and ray-bilinear patch intersections. Our method is faster than comparable approaches for ray tracing, enabling real-time surface editing.
\end{abstract}

% keywords can be removed

\section{Introduction}
Displacement maps enable artists to model complex features over three-dimensional surfaces. Authoring displacement maps is a natural way to add detail to base meshes, especially when limited resources preclude access to or compression of highly detailed source assets. Therefore, we focus on the look development of depth-based surfaces to be applied to arbitrary models. The examination of details through reflections, refractions and shadows provides valuable visual feedback, where the ideal authoring system would perform ray tracing during editing. Since ray tracing performance is directly tied to interactivity, we explore authoring as both an editing and a ray tracing performance problem. The additional constraint of real-time editability changes the range of ray tracing solutions considered. 

Very early editing systems, such as 3D Paint \cite{Williams1990}, allow for view-dependent interactive editing of displacement maps with non-real time updates. Distance functions simplify editing with trivial operations, yet these volumetric functions have special storage and rendering requirements \cite{Peng2004}\cite{Reiner2011}. Specialized surfaces, such as terrain, can be easier to manipulate \cite{Yusov2012}. Editing interactions may also be limited, for example, to only penetrating collisions with relief mapping  \cite{Nykl2014}, or to cut-and-paste with subdivision surfaces \cite{Biermann2002}. Subdivision surfaces have received considerable attention \cite{Zorin1997}\cite{Kobbelt1998}\cite{Schaefer2015}\cite{Brainerd2016}, although artists must be aware of the subdivision workflow. We explore the interactive editing of high-resolution displaced surface details over arbitrary low-poly base meshes with real-time ray tracing. 

We note that interactive editing is primarily a performance issue, as both rendering and data structure modifications must be computed in real-time. Therefore, we consider editability as a ray tracing performance problem with the additional constraint that all data structure updates must also occur in real-time. Many techniques for ray tracing detailed surfaces have been developed, including direct sampling of depth textures via parallax, relief or cone mapping  \cite{Blinn1978}\cite{Kaneko2001}\cite{Oliveira2000}, ray tracing of parametric intervals  \cite{Lischinski1990}\cite{Wang2002}\cite{Thonat2021}, and tessellation techniques \cite{Snyder1987}\cite{JangHan2013}\cite{Niessner2013}. A recent trend involves the use of locally adaptive tessellation to render massive static scenes  \cite{Karis2021}\cite{Benthin2023}\cite{Haydel2023}. Since we are not concerned with real-time rendering of large static scenes but with the look development of specific assets, we avoid solutions that require dynamic BVH updates or complex pre-computation. Instead, we favor of direct sampling methods that offer both performance and quality. Our tessellation-free method renders detailed, displacement-mapped surfaces with interactive editing and ray tracing.

\begin{figure}
  \includegraphics[width=\textwidth]{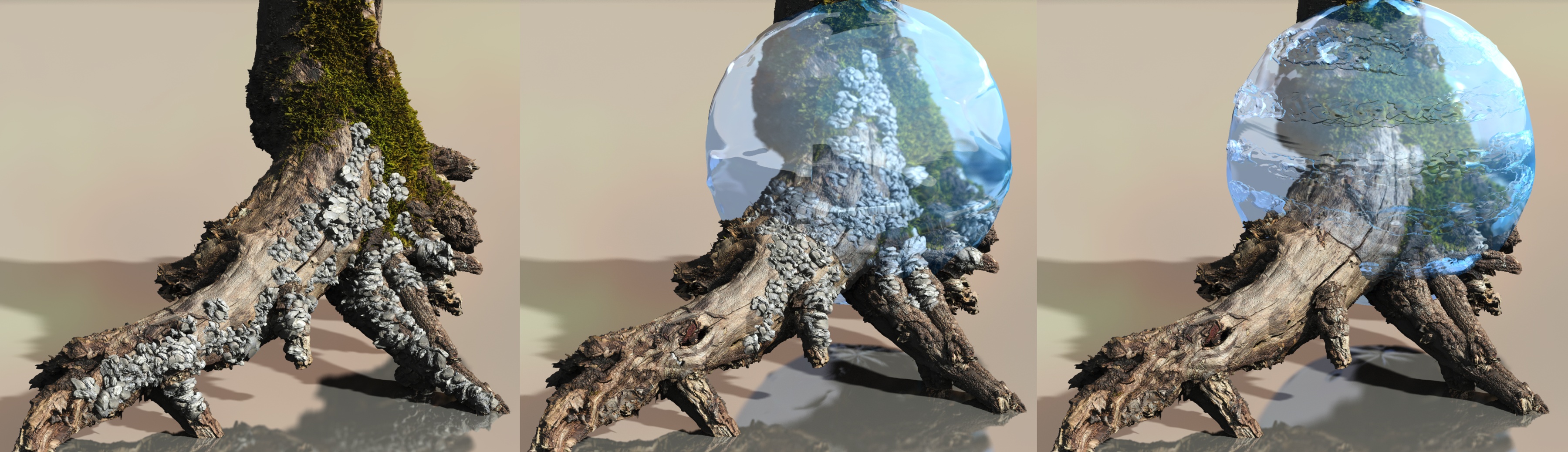}
  \caption{Detailed geometry is interactively edited and ray traced with our technique. A base mesh of 7710 triangles with a 16-bit, $4096^2$ displacement map is ray traced at 3460x1024 in 28 milliseconds for one sample including primary rays, path tracing, reflection, refraction and shadows. Interactive sculpting of the displacement map while ray tracing supports a) sculpting rocks over a tree stump, b) sculpting while looking through a refractive object, and c) modeling the surface of the refractive object itself.}
  \label{fig:teaser}
\end{figure}

\section{Related Work}

\textit{Parallax mapping} renders depth details by offsetting texture samples based on parallax (shifted occlusion) \cite{Kaneko2001}. \textit{Parallax occlusion mapping} iteratively samples the height field to ensure correct depth while also supporting occlusion shadows \cite{Tatarchuck2005}. \textit{Prism parallax occlusion} mapping enhances this with correct silhouettes by intersecting view rays with extruded prisms split into three tetrahedra \cite{Dachsbacher2007}. \textit{Relief maps} encode the displaced surface as a color texture with an additional depth channel. The original method was an image-warping technique  \cite{Oliveira2000}, whereas later methods traverse the relief map to cover arbitrary base meshes \cite{Policarpo2007}. Quadric surfaces may also be covered by relief maps by using ray-quadric intersection before relief stepping \cite{Oliveira2005}. Intended for raster pipelines, parallax and relief mapping do not utilize BVHs or support general ray tracing. 

\textit{Cone step mapping} accelerates the intersection search by precomputing a cone map to skip features based on occlusion angle  \cite{Donnelly2005}\cite{Dummer2006}, employing binary search \cite{Policarpo2007}, or using a min-max texture \cite{Lee2009}. However, anti-aliasing cone maps introduces overhead since they cannot be hardware interpolated. These methods are view-dependent, making them unsuitable for ray tracing arbitrary base meshes. 

\textit{Curved rays} map the 3D volume of the displaced surface to nonlinear texture space \cite{Kajiya1983}. Tangent-space methods compute the curved mapping between world and depth texture space at each sample, approximating curved rays with piecewise segments \cite{Chen2008}. We avoid this transform per sample by designing a parallel offset prism with a linear projection to texture space. \cite{Ogaki2023} develops nonlinear rays by solving cubic equations that map world rays into canonical prism volumes. In their work, multiple data structures are provided for acceleration: BVHs for broad phase, and min-max mipmaps for local displacement or BVHs for instanced meso-geometry. Rays are transformed at the leaf nodes (prisms) of the broad-phase BVH, where they become nonlinear in rectified canonical space (see Fig. \ref{fig:prisms}, top right). Although optimizations are mentioned, the pre-computation of data structures and solving cubic equations preclude our goals for interactive editing.

\textit{Parametric surfaces} may be exactly evaluated through root-finding with Newton’s method, whereas complex surfaces must be divided into multiple intervals \cite{Toth1985}\cite{Joy1986}. To guarantee convergence, the intervals must be chosen to locally bound the surface at minima and maxima. \cite{Lischinski1990} construct a tree structure to efficiently traverse the surface. Caching surface trees enables fast parallel ray tracing of parametric surfaces by sharing information with neighboring rays \cite{Lamotte1991}. 

\textit{Affine intervals} have been shown to provide better local bounds than min-max intervals \cite{Knoll2009}. \cite{Thonat2021} construct a D-BVH as a min-max mipmap over affine intervals of the base geometry. Displaced bounds are generated dynamically from quad-tree nodes, requiring significant computation for intersection. Recent work improves performance with a hierarchical data structure, RMIP, over the displacement map using a bidirectional mapping between 2D depth texture and 3D object space  \cite{Thonat2023}. Rectangular regions in RMIP allow for acceleration of anisotropic ray sampling. Since D-BVH and RMIP can be costly during traversal, we explore a direct sampling approach for quality and performance during editing. 
 
\textit{Shell mapping} constructs a bijective map between texture space and shell space - the region between two offset surfaces - where meso-geometry will be rendered locally \cite{Porumbescu2005}. Rays are traced in shell space and transformed back to world space, mapping a 3D volume onto a surface. Real-time rendering is achieved by \cite{Ritsche2006} utilizing a distance map to accelerate ray tracing within the shell space of the primary object and a volumetric texture to encode the meso-geometry. Both techniques precompute the shell space prism as three tetrahedra.  \textit{Curved shell mapping} removes discontinuities by modeling Coons patches within each prism at the cost of solving a cubic equation per ray step \cite{Jeschke2007}.

A different approach to meso-geometry finds ways to precompute intersections through suitable encodings. One such encoding uses singular value decomposition \cite{Wang2003}, while others use neural networks to evaluate intersections \cite{Kuznetsov2021}. Their work handles meso-geometries yet encoding requires considerable pre-processing that we avoid in favor of interactive performance and editability.

\textit{Tessellation} is an intuitive approach to displacement mapping, generating additional geometry for detail as needed  \cite{Cook1984}. \cite{Snyder1987} accelerate fully tessellated surfaces with uniform spatial grids. Full tessellation is memory-intensive, thus research shifted to localized, micro-tessellation techniques, such as geometry caching of sub-triangles \cite{Pharr1996}. \textit{Adaptive hardware tessellation} was introduced as a shader pipeline stage with DirectX 11 \cite{Mic2009}  enabling real-time adaptive displacement and higher-order surfaces \cite{Niessner2016}.  The tessellation stage can be used to adaptively render displacement-mapped surfaces \cite{JangHan2012}\cite{JangHan2013}. However, direct hardware tessellation is generally poorly suited to ray tracing since non-view facing reflections and refractions are not easily handled.

\textit{Higher order surfaces} are a commonly tessellated, including Bézier patches  \cite{Munkberg2010}\cite{Concheiro2011}\cite{Tejima2015}, Catmull-Clark surfaces \cite{Niessner2013}, Gregory patches \cite{Loop2009}, and Loop subdivision surfaces \cite{StamLoop2003}\cite{Lee2000}\cite{Amresh2012}. These surfaces provide smoothness guarantees with $C^1$ or $C^2$ continuity. We introduce a novel adjustment to the displaced shading normal that achieves visual continuity when ray tracing arbitrary base meshes without an intermediate $C^1$ surface.

\textit{Locally adaptive tessellation} is a recent trend whereby GPU tessellation is applied to increase detail locally on arbitrary base meshes. Nanite introduced virtual geometry to compress source geometry into small triangle clusters which are loaded and rendered as dynamic, regional levels of detail (LOD) on-the-fly \cite{Karis2021}. To support ray tracing, \cite{Benthin2023} insert triangle clusters into a dynamic, GPU-accelerated BVH per frame. These methods depend on a high-resolution source mesh which is preprocessed through clustering and decimation of triangle groups.  \cite{Haydel2023} introduces locally adaptive ray tracing which does not split the mesh but instead refines each triangle through an explicit one-to-four triangle subdivision. Pre-computed acceleration is achieved by inserting each division into a custom tessellation tree, which can be used to select per-triangle LODs on-the-fly.

Micro-Meshes \cite{Nvidia2023} is an API for rendering highly detailed objects as locally adaptive subdivided triangles with data compression via barycentric maps. This eliminates the need for UV mapping since sampling is implicit. Primarily  intended to compress detailed base models, displacement mapping is supported by remapping from a depth texture to the barycentric map. Although this precludes interactive editing we examine Micro-Meshes in terms of rendering quality compared to our approach. 

Modern techniques for locally adaptive tessellation can give good rendering performance over highly detailed source meshes, yet this is often at the cost of complex data structures utilizing the most recent GPU hardware.

\begin{figure}
  \includegraphics[width=\textwidth]{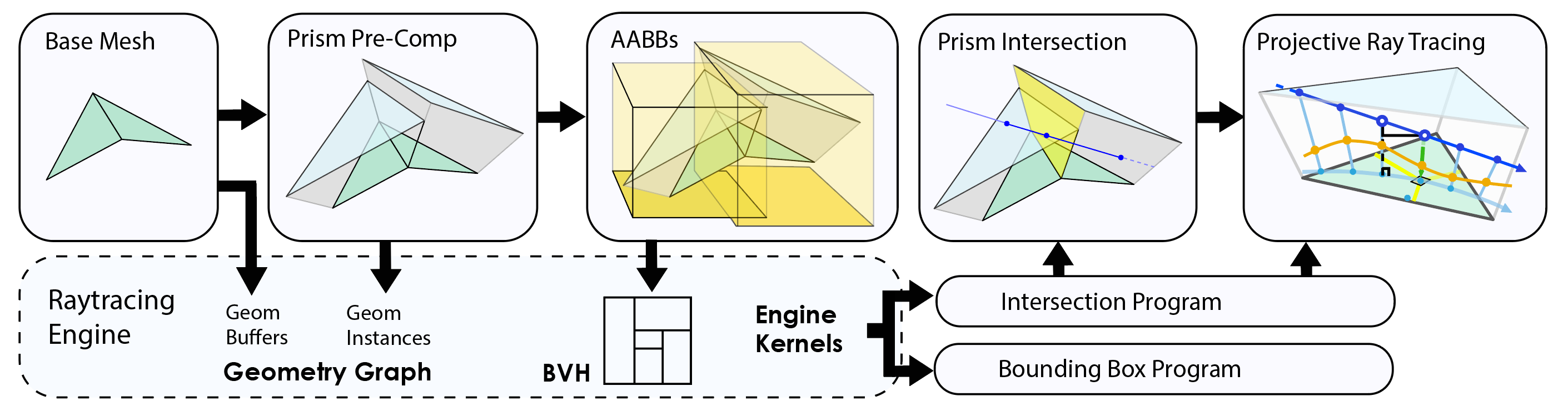}
  \caption{\textbf{Method overview}. Our technique starts with a low-poly base mesh, from which we pre-compute parallel offset prisms with positive and negative extents. Prisms and their AABBs are used to build the geometry graph and hardware-accelerated BVH. A custom program implements prism intersection and the Projective Displacement algorithm described in text.}
 \label{fig:overview}
\end{figure}
 
\section{Overview}

\subsection{Motivation}

Our goal is to enable the interactive editing of complex details over meshes while performing real-time ray tracing. We focus on the look development of single objects, excluding the massive rendering of static scenes. Recognizing that many studios or independent artists may lack access to cutting-edge hardware, extensive asset libraries, or sufficient GPU memory, we deliberately avoid approaches that require compression or streaming of highly detailed source assets or frequent data structure rebuilding. Instead, our method supports the authoring of displacement-mapped surfaces with a minimal memory footprint, relying on a low-poly mesh and a high-resolution depth texture, while addressing the added performance constraints of real-time editing.
 
\subsection{Algorithm Design}

Modern locally-adaptive ray tracing techniques distinguish between top-level acceleration structures (TLAS), for high level objects, and bottom-level acceleration structures (BLAS) for procedural primitives and surface details \cite{Wyman2018}\cite{Sanzharov2019}\cite{Kacerik2023}. We leverage existing hardware BVHs for TLAS to support base objects with arbitrary meshes composed of triangular faces extruded into prisms. While displacement acceleration options for bottom-level primitive BLAS may include mipmaps  \cite{Thonat2021}\cite{Thonat2023} and dynamic BVHs \cite{Benthin2023}\cite{Ogaki2023}, the requirement to regenerate data structures when details are edited present a mismatch with our goals. Complexity and dynamic surface changes increase overhead in BLAS, especially when the real-time frame rate must include reconstruction and local per-ray data structure traversals. We, therefore, avoid a BLAS and improve direct sampling methods for prismatic primitives. Our solution may be considered as a novel direct sampling method integrated with a hardware-based BVH for TLAS in ray tracing. 

\begin{figure}
  \includegraphics[width=\textwidth]{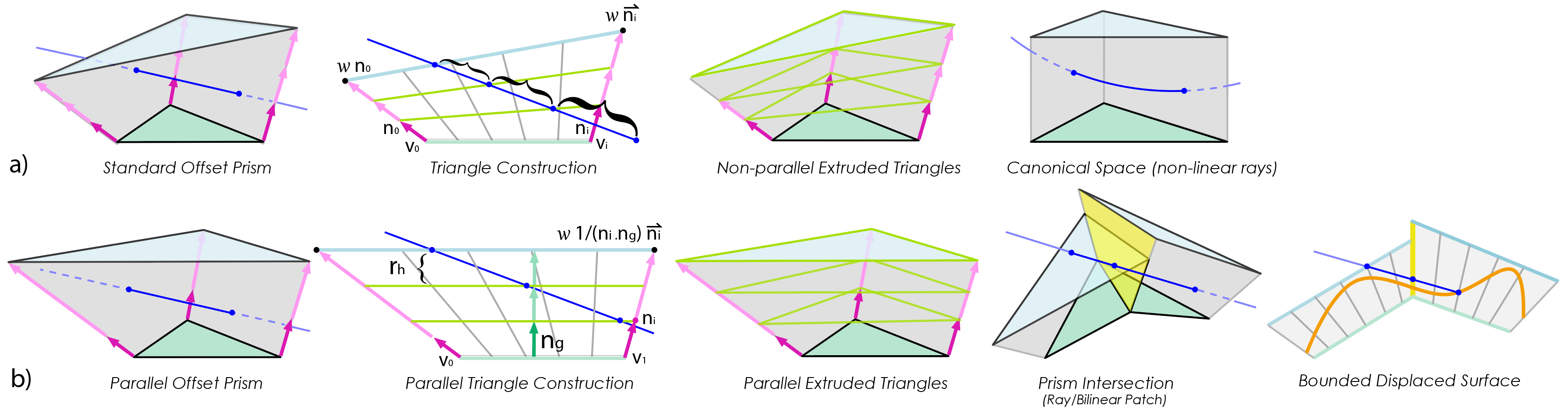}
  \caption{\textbf{Prism construction}. a) Typical offset construction is based on multiples of the vertex normals which results in non-parallel extruded triangles and a non-linear sample space (brackets). One approach to handle this is by converting to canonical space in which rays are non-linear. b) Our method constructs a prism volume with parallel offsets by adjusting the normal length relative to the geometric normal resulting in parallel extruded triangles and a linear sample space. Rays intersect a prism at either triangles or bilinear patches.} 
  \label{fig:prisms}
 \end{figure}

Efficient ray marching is challenging since equally spaced world samples are non-linear in texture space due to the typical offset construction of prisms (Fig. \ref{fig:prisms}a). Among the previous  direct sampling methods only Prism Parallax Occlusion mapping (PPOM) produces correct silhouettes - and achieves this with offset prisms - yet has difficulty with ambient occlusion in the raster pipeline \cite{Dachsbacher2007}. We construct parallel offset prisms such that world-space samples can be \textit{linearly projected} to texture space (Fig. \ref{fig:prisms}b).

Our algorithm introduces several novel techniques for integrating projective direct sampling with a top-level hardware-accelerated BVH. These contributions include:

\begin{itemize}
\item Parallel offset prisms from a base mesh where ray samples can be linearly projected to texture space
\item A new projective displacement mapping (PDM) method that efficiently ray marches to find the intersection depth within a parallel offset prism
\item A top-level hardware BVH with prism entry accelerated by ray-bilinear patch intersections, eliminating the need for tetrahedra
\item Optimized entry/exit conditions with watertightness
\item A novel method for normal correction that retains displaced features while avoiding $C^1$ intermediate surfaces even with very low poly base meshes
\item Overall simplicity, with no BLAS, leveraging hardware BVHs for TLAS
\end{itemize}

Projective displacement mapping is a local sampling method that directly accesses the displacement map, whereas our overall technique relies on hardware BVH structures in Nvidia OptiX for the TLAS of the base mesh. We design bounding box and intersection programs that perform projective displacement sampling. Although editing the base mesh requires a BVH rebuild, direct editing of the displacement map is possible without data structure rebuilds.

\section{Prism Construction}

Prisms typically used to define displacement mapping are constructed by offsetting base triangles along their vertex normals. Given a base triangle in barycentric coordinates defined by $P(u,v) \in \mathbb{R}^3$, with $0 \leq u,v \leq 1$, a standard offset prism $\textbf{R}_{std}$ with vertices $V_i$ and normals $N_i$ is given by:

\begin{align}
\mathbf{P}(u,v) &= u V_0 + v V_1 + (1-u-v) V_2, \\
\mathbf{N}'(u,v) &= u N_0 + v N_1 + (1-u-v) N_2, \\
\mathbf{R}_{std}(u,v,w) &= \mathbf{P}(u,v) + w \mathbf{N}'(u,v)
\end{align}

The maximum extent is defined by a constant $w_{max}$ and the offset triangle $\textbf{P}(u,v,w_{max})$. In Figure \ref{fig:prisms}a, note that offset triangles in standard prisms may not be parallel to the base triangle since the unit normals point in different directions. As observed by others the projection of a ray in this volume to texture space is non-linear. Some ray tracing methods transform this irregular prism to canonical space where the rays are nonlinear \cite{Chen2008}\cite{Ogaki2023}, Figure \ref{fig:prisms}a (top right).

We introduce a \textit{parallel offset prism}, $\textbf{R}_{pop}$, whose normal directions are identical, yet whose offset triangles are all parallel to the base triangle, Figure \ref{fig:prisms}b.

\begin{align}
n_{f}(u,v) &= u \frac{1}{N_0 \cdot N_g} + v \frac{1}{N_1 \cdot N_g} + (1-u-v) \frac{1}{N_2 \cdot N_g} \\
\mathbf{R}_{pop}(u,v,w) &= \mathbf{P}(u,v) + w n_{f}(u,v) \mathbf{N}'(u,v)
\end{align}

The normal factor $n_{f}(u,v)$ transforms the space to an orthogonal one where $w$ can now be interpreted as distance along the geometric normal $N_g$. Note these triangles are parallel but not similar triangles, since the arbitrary normal directions still shift the vertices laterally. 

Maximum prism extents $e_i$ are computed at the vertices as:

\begin{equation}
\label{eqn:01}
e_i = v_i + w_{max} \frac{1}{N_i \cdot N_g} N_i
\end{equation}

A displaced surface $\textbf{S}$ is defined by a depth function $D(u,v)$ over the base triangle as:
\begin{equation}
\label{eqn:surface}
\mathbf{S}(u,v) = \mathbf{P}(u,v) + D(u,v) \mathbf{N}'(u, v)
\end{equation}

\begin{figure*}
  \includegraphics[width=\textwidth]{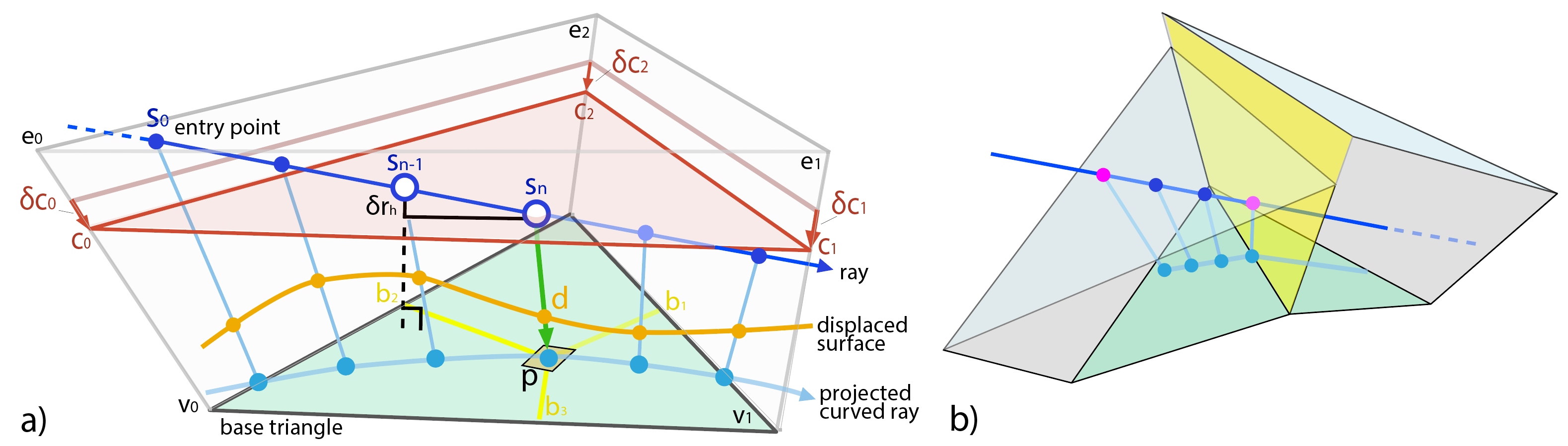}
  \caption{\textbf{Projective displacement mapping}. a) Surface intersection proceeds along a ray while simultaneously tracking a parallel offset triangle (scanning triangle), $c_i$, where the barycentric projection of a sample $s_n$ gives the displacement texture location $d$. See text for details. b) When a ray enters or exits a prism it intersects at either a triangle top/bottom or a bilinear patch. Prism watertightness is achieved by ensuring sample continuity across these boundaries.}
  \label{fig:pdm}
\end{figure*}

$\mathbf{S}$ is constructed from $\mathbf{N}'$, the interpolated (shading) normal at each point. This is the most common formulation for a displaced surface \cite{Cook1984}\cite{Lee2000}\cite{Niessner2013}. We observe that $D(u,v)$ replaces the parametric scalars in Eqn. 3 and 4. This implies that any bounding prism constructed as a scalar multiple along the same normals will result in identical displaced surfaces, in this case so long as $D(u,v) < w_{max} n_{f}(u,v)$.

More importantly, changes in ray height ($r_h$ in Fig. \ref{fig:pdm}b) for a given sample's corresponding offset triangle (green lines) can both be evaluated linearly along $N_g$. This enables our novel \textit{projective displacement mapping} (PDM) technique which shares similarities to both direct sampling and curved rays, yet is considerably faster. The PDM algorithm and its optimizations are described in section 4.3.

\subsection{Prism Intersection}

The interface between two prisms is a bilinear patch since adjacent vertex normals $v_i$ may not be co-planar \cite{Ramsey2004}\cite{Reshetov2019}. \cite{Jeschke2007} demonstrate that using interpolated normals can construct prisms with $C^1$ continuity internally and $C^0$ continuity for adjacent prisms. To handle the interface, others have consider tetrahedra, deconstructing the side faces into two triangles each, where the diagonal must be consistently selected to avoid artifacts \cite{Hirche2004}\cite{Dachsbacher2007}. We take a different approach and directly ray-trace the prism sides as bilinear patches.

Direct ray tracing of bilinear patches has several benefits. First, the interface between two sides can support $C^1$ continuity since it is not being approximated by two planar triangles. Second, there is no logic or storage needed for consistent diagonal selection. Third, rather than tracing the six exterior faces we can trace just three bilinear patch primitives directly. Finally, recent results provide efficient raytracing of bilinear patches on GPU hardware \cite{Reshetov2019}. The loop condition of our primary algorithm requires both entry and exit t-values which can be computed more quickly together. Given ${n_p}$ as the normal at the patch hit point:

\begin{equation}
\begin{split}
t_{max} & = max( t, t_{max}) \textbf{ iff } n_p \cdot r_{dir} \geq 0 \\
t_{min} & = min( t, t_{min}) \textbf{ iff } n_p \cdot r_{dir} < 0
\end{split}
\end{equation}

Rays may enter a bilinear patch side, through the top triangle of a prism, or start inside the prism. Top faces are handled with an extra ray/triangle test. We can support rays which start inside the prism by observing when $t_{max}$ is set but $t_{min}$ is not. These values are updated similarly for the three bilinear patch sides and the top and bottom triangles.

% Algorithm
\begin{algorithm}[t]
\SetAlgoNoLine
\KwIn{Prism: negative extents $v_i$, positive extents $e_i$, vertex normals $n_i$, Ray: $r_0$,$r_d$, and Displacement Map: $f$
}
\KwOut{Hit t-value: $t$, Hit point: $s$, and Normal $n$}
\vspace{0.2cm}  
\If{ intersectPrism( $v_i$, $e_i$, $t_{min}$, $t_{max}$)\hspace{0.1cm} }
{
  $t = t_{min}$ \\
  $s = r_0 + r_{dir} t$  \hspace{\fill}initial hit point\\
  $h_{ray} = \textit{pointToPlane}( s, v_i )$ \hspace{\fill}initial sample height \\
  $c_i = v_i + (e_i - v_i) h_{ray}$ \hspace{\fill}initial scanning triangle\\
  \vspace{0.2cm}  
  \For{ $h_{ray} > h_{surf}$ \textit{ and } $t < t_{max}$ }
  {  
     $\delta h = n_s \cdot (c_0 - s)$  \hspace{\fill}point-to-plane distance\\
     $c_i = c_i - n_i \delta h$ \hspace{\fill}advance scanning triangle\\
     $b_i = \textit{triangleBarycentric}( s, c_i )$ \hspace{\fill}barycentric coords\\
     $p = \vec{v}_0 b_0 + \vec{v}_1 b_1 + \vec{v}_2 b_2$ \hspace{\fill}projected point\\
     $h_{ray} = |s - p|$ \hspace{\fill}ray sample height\\
     $uv = \vec{uv}_0 b_0 + \vec{uv}_1 b_1 + \vec{uv}_2 b_2$\hspace{\fill}uv coords\\
     $h_{surf} = \textit{textureSample}( f, uv )$ \hspace{\fill}surface height\\     
     $t = t + dt$\\
     $s = s + r_{dir} dt$\\
  }
  \vspace{0.2cm}  
  \If{ $h_{ray} < h_{surf}$ }
  {
     $t = \textit{interpolateCrossing}( t, dt, h_{surf} )$ \hspace{\fill}final hit\\
     $p_{hit} = r_0 + r_{dir} t$ \\
     $b_i = \textit{triangleBarycentric}( p_{hit}, c_i )$ \hspace{\fill}final barycentric coords\\
     $N' = \vec{n}_0 b_0 + \vec{n}_1 b_1 + \vec{n}_2 b_2$ \hspace{\fill}interpolated base normal\\
     $uv_a = \vec{uv}_0 b_0 + \vec{uv}_1 b_1 + \vec{uv}_2 b_2$ \hspace{\fill}finite difference uvs\\
     $uv_b = \vec{uv}_0 (b_0 + \delta b_x) + \vec{uv}_1 b_1 + \vec{uv}_2 (b_2 - \delta b_x)$\\
     $uv_c = \vec{uv}_0 b_0 + \vec{uv}_1 (b_1 + \delta b_y) + \vec{uv}_2 (b_2 - \delta b_y)$\\
     $p_a = \vec{v}_0 b_0 + \vec{v}_1 b_1 + \vec{n}_2 b_2$ \hspace{\fill} base points\\
     $p_b = \vec{v}_0 (b_0 + \delta b_x) + \vec{v}_1 b_1 + \vec{n}_2 (b_2 - \delta b_x)$\\
     $p_c = \vec{v}_0 b_0 + \vec{v}_1 (b_1 + \delta b_y) + \vec{n}_2 (b_2 - \delta b_y)$\\
     $s_a = p_a + N' \textit{textureSample}( f, uv_a )$\\
     $s_b = p_b + N' \textit{textureSample}( f, uv_b )$     \hspace{\fill}surface tangent\\           	     	     
     $s_c = p_c + N' \textit{textureSample}( f, uv_c )$ \hspace{\fill}and bi-tangent\\     
     $N_s = (s_c-s_a) \times (s_b-s_a)$ \hspace{\fill}displaced surface normal\\
     $N'_s = N_s - N_g + N'$ \hspace{\fill}corrected normal (Sec 5.1)\\
     \textbf{return } $t, p_{hit}, N'_s$\\
  }
}
\caption{Projective Displacement Mapping Program}
\label{alg:one}
\end{algorithm}

\subsection{Raytracing with Projective Displacement}

The key idea of projective displacement mapping (PDM) is to sample the ray linearly in world space and then map this to non-linear barycentric coordinates while avoiding costly tangent-space matrix transforms. Starting with a prism defined by negative extents $v_i$ and positive extents $e_i$, and ray $r_0$,$r_d$, our technique enters the prism at a bilinear patch or triangle face with entry point $s_0$, see Figure \ref{fig:pdm}a. A parallel offset triangle, $c_i$, is constructed based on the height of the entry point and then advanced along the delta vectors $\delta c_i$ according to the perpendicular height $\delta r_h$ of the previous sample $s_{n-1}$. 

From the scanning triangle $c$ and point $s_n$ we can compute barycentric coordinates $b_i$ which represent the projection of $s_n$ along the interpolated normal to the base triangle point $p$, proof of which is provided in Appendix A. These coordinates are used to derive UV coordinates to fetch a displacement texel to compute the surface height $h_{surf} = |d-p|$, whose comparison to the ray sample height $h_{ray} = |s-p|$ determines if the surface has been hit.

The main loop is optimized by avoiding boundary checks across each prism face. Instead, we leverage the earlier prism intersections which provide $t_{min}$ and $t_{max}$ as simplified loop bounds. Additionally, the scanning triangle is updated by computing the incremental change in height $\delta h$ that moves the previous triangle to the next sample location. Note that we cannot simply update the ray height $h_{ray}$ with $\delta r_{h}$ because the sampling heights $h_{ray}$ and $h_{surf}$ are measured along the interpolated normal whereas $\delta r_{h}$ is a change in height measured along the geometric normal $n_g$ of the base triangle.

The sequence of steps is illustrated in Figure \ref{fig:pdm}, with the algorithm provided in Listing 1. The function \textit{triangleBarycentric} computes the barycentric coordinates of the sample point relative to the scanning triangle with five dot products \cite{Ericson2004}. Practically, the \textit{textureSample} function evaluates the displacement map with hardware bilinear interpolation with offset and scaling so that the [0,1] input range is translated into a world space height relative to the base triangle $v_i$. Once a hit is detected an interpolation is performed across the surface boundary to resolve a more accurate hit. Finally, the displaced surface normal is computed by taking differentials in UV space and reconstructing the tangent and bitangent at the intersection point.

 \begin{figure}
  \includegraphics[width=\textwidth]{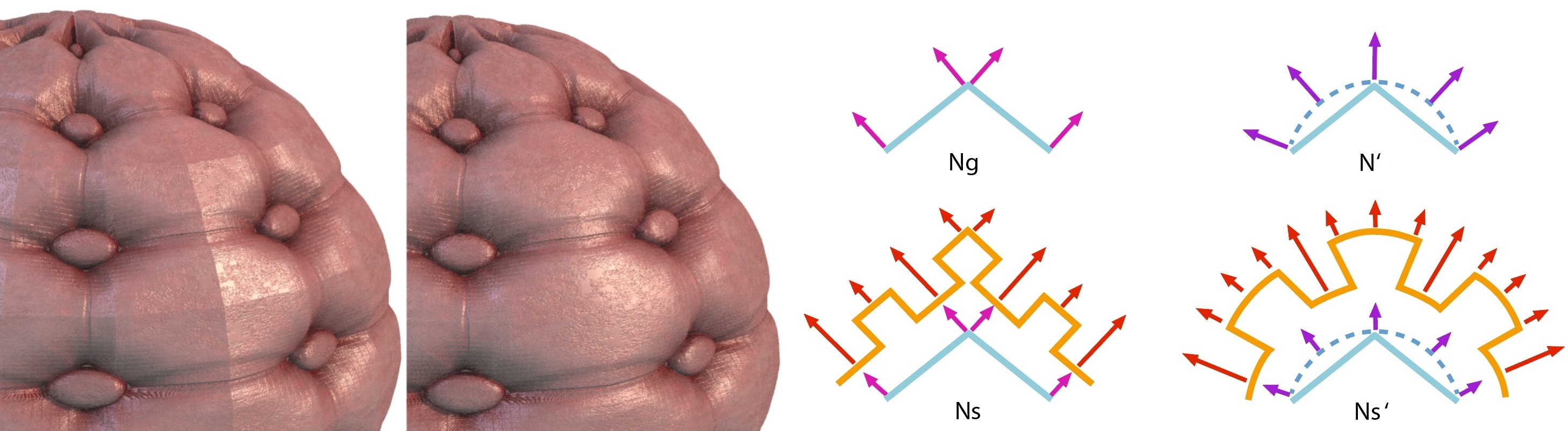}
  \caption{\textbf{Shading normal correction}. In low polygon meshes, when the angle between base triangles is high and no intermediate $C^1$ surface is used, the geometric normal $N_g$ can be noticeably carried through to the displaced surface $N_s$. This can be corrected to give smoothed displaced normals $N'_s$ by subtracting the geometric normal and adding in the interpolated normal $N'$, as discussed in Section 5.1.}
  \label{fig:shading}
 \end{figure}
  
\section{Smoothness, Quality and Details}

\subsection{Shading Normal Continuity}
In several works on displacement mapping, the authors make use of intermediate surfaces such as Catmull-Clark or Bezier surfaces to achieve $C^1$ continuity across base triangle boundaries \cite{Smits2000}\cite{Lee2000}\cite{Jeschke2007}\cite{Niessner2013}. Discontinuities may be visible at base mesh boundaries if this is not done. We quantify and correct for these artifacts for the sake of efficiency without resorting to $C^1$ surfaces.  

Shading discontinuities are observed when the angle between base triangles is high and the displacement map is relatively smooth. An example is provided in Figure \ref{fig:shading}. We prove in Appendix B that the displaced surface normal $N_s$, evaluated numerically at the hit point $p$ with finite differences, contains the flat shading normal of a base triangle $N_g$, even though the surface $\mathbf{S}$ is computed from the interpolated normal $\mathbf{N}'$ in Eqn. 7. That is:

\begin{equation}
    \mathbf{N}_s = \mathbf{N}_g + \nabla D(u,v) \otimes \nabla_{\mathbf{P}} \mathbf{N}'(u,v)
\end{equation}

The term $N_g$ introduces boundary artifacts across triangles in the base mesh, shown in Fig. \ref{fig:shading} (left) and Fig. \ref{fig:visual1} (MicroMesh and RMIP). Similar to Phong shading \cite{Phong1973}, we wish to replace the geometric normal $N_g$ with the interpolated base normal $N'$. This can be done directly as:

\begin{equation}
 \mathbf{N}'_s = \mathbf{N}_s - \mathbf{N}_g + \mathbf{N}'
\end{equation}
where $N_s$ is the displaced normal computed in Algorithm 1 and $N'_s$ is the corrected normal. This retains the surface features of displacement while improving shading continuity. The resulting surface still has micro-bumps, displacement shadows and correct silhouettes.

\subsection{Watertightness}
Watertightness in our method may refer to \textit{geometric watertightness} at prism boundaries or to \textit{surface watertightness} across the displaced surface itself. Since there is no tessellation the only geometric boundaries are between prisms. The mapping from prisms to texture space is non-linear, where the interface between prisms is a bilinear patch. With shell mapping, Porumbescu et al. noted that piecewise approximations of this interface with tetrahedra result in "buckling" artifacts \cite{Porumbescu2005}. We avoid this by directly raytracing prism interfaces as bilinear patches. Geometric watertightness at prism boundaries is therefore as good as the ray/bilinear patch algorithm "cool patches" we use from \cite{Reshetov2019}.

Surface watertightness considers whether sampling is guaranteed to find the displaced surface when it exists along the ray. \cite{Ogaki2023} found such issues with non-linear ray tracing and reduced them with double precision. Surface holes are avoided in our method so long as the minimum displacement height is larger than the sample spacing $dt$, that is $min( D(u,v) ) > dt$. While no thru-holes will occur, thin features are addressed in the next section (5.3).

\subsection{Thin Features}
Ray marching algorithms are known to miss features thinner than $dt$. We address this by stochastically shifting samples along $t$ so that such features are integrated over multiple sample frames. Typically used to distribute rays across pixels, light sources, or surface BRDFs, the ray samples in a prism are shifted by $t_0 = t_{hit} + R dt $ with a random number $R \sim \mathcal{U}(0, 1)$. This ensures that a sample sequence covers thin features over many sample frames. The first ray sample $s_0$ must be jittered \textit{after} prism intersection and entry. See Figure \ref{fig:visual2} for comparisons with and without thin feature sampling.

\begin{figure}[h]
  \centering
  \includegraphics[width=0.7\textwidth]{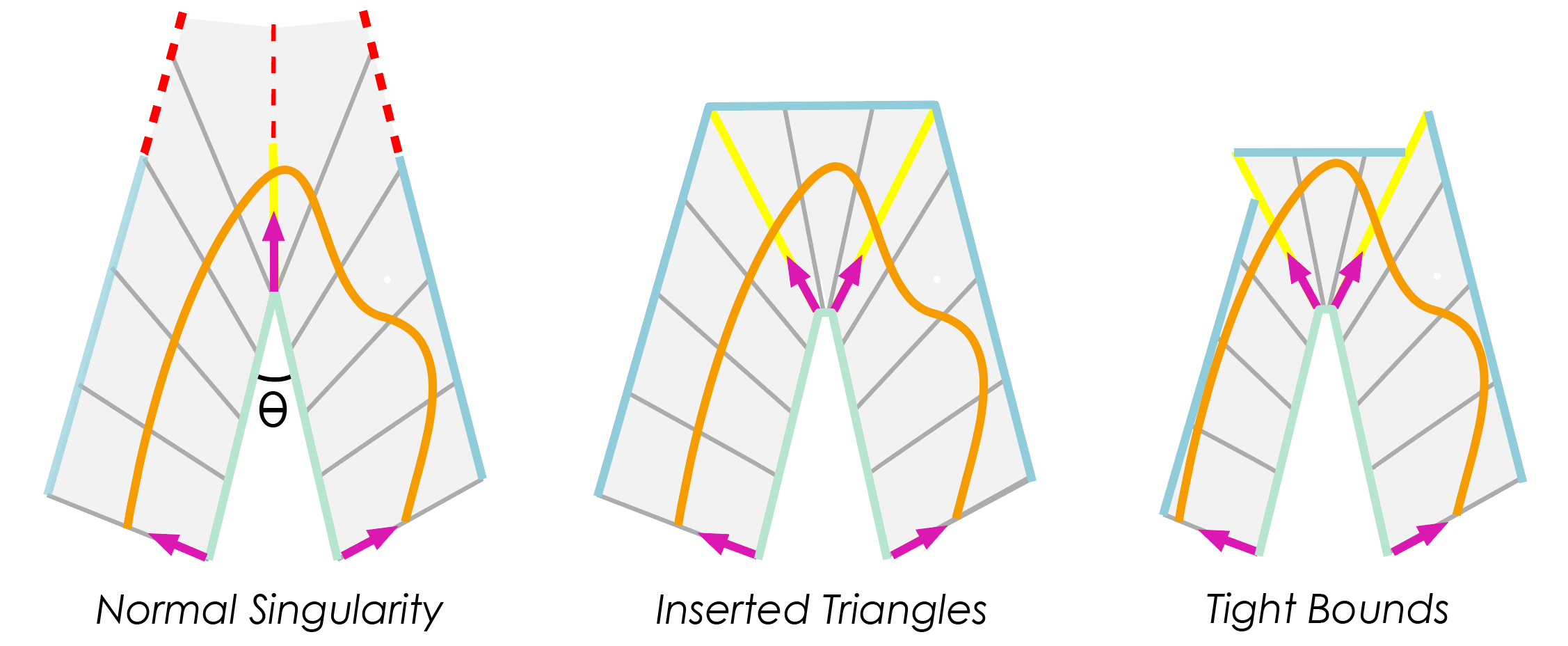}
  \caption{\textbf{Offset distance}. Sharp crease edges can cause singularities (left), which are corrected by inserting base triangles at these edges. Tighter bounds can be achieved by setting maximum offset extents per prism (right).}
  \label{fig:offset}
\end{figure} 

\subsection{Offset Distance}
Care must be taken to ensure that the maximum offset distance of prisms will bound the displaced surface. If two triangles meet at a sharp crease edge (Fig \ref{fig:offset}a), the denominator $N_i \cdot N_g$ in Equation 6 causes a singularity and the normal extension goes to infinity. A solution is to insert two new triangles at any such edges whose interior angle is less than some threshold (eg. $\theta < 5^{\circ}$). This issue occurred in fewer than 10 triangles in two of the models we used (Stanford Dragon and Tree Stump). 
 
The offset distance, $w_{max} \frac{1}{N_i \cdot N_g}$, determines the maximum displacement from the base mesh. So long as this is everywhere greater than D(u,v) the displaced surface will be fully enclosed by the shell region. We can ensure tighter bounds and faster performance, by assigning $w_{max}$ per prism at the cost of pre-computing $max(D(u,v))$ for each triangle over the depth texture. Although this can result in a misalignment in the top faces between two prisms (Fig 6c), the height of the displaced surface does not depend on the bounding prism extents and the surface is still found below both of these.
 
\begin{figure}
  \includegraphics[width=\textwidth]{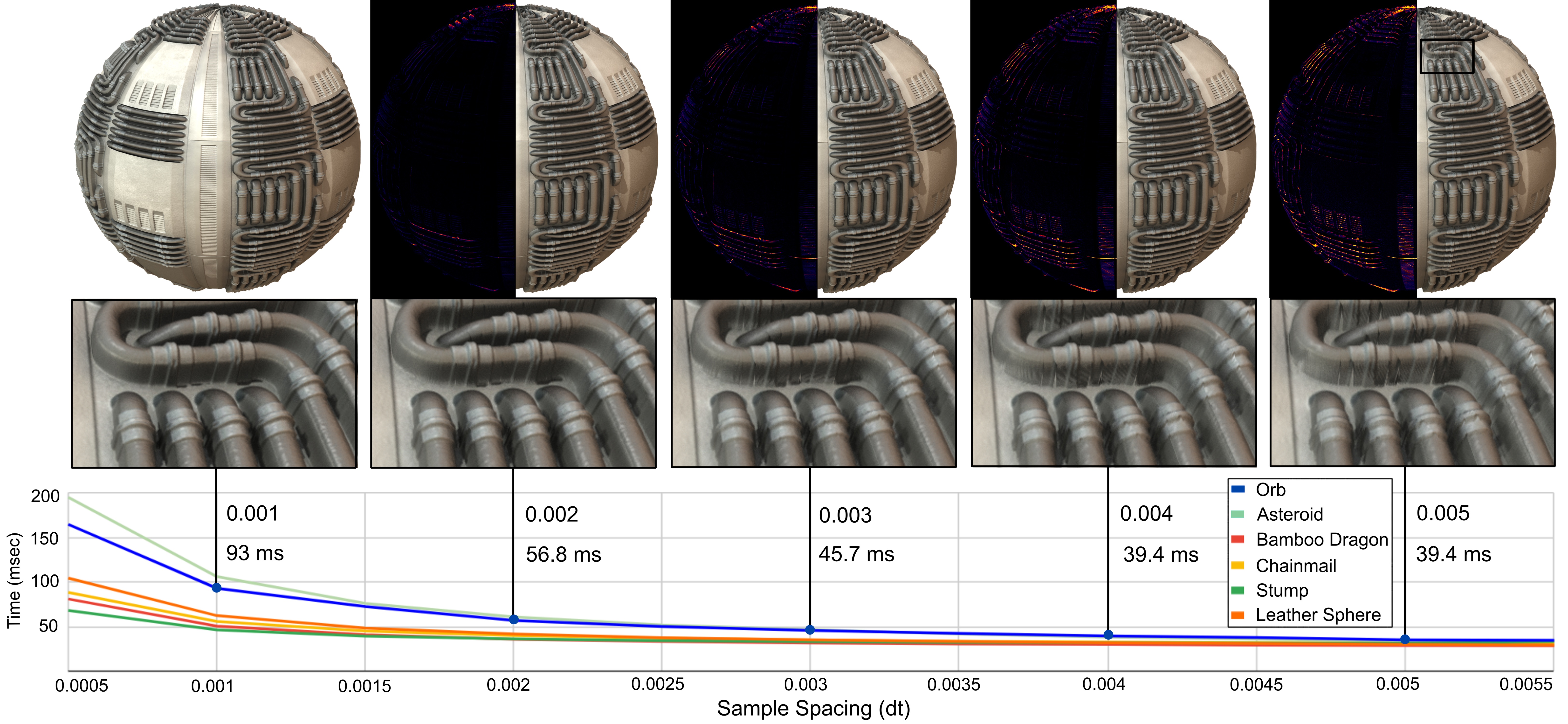}
  \caption{\textbf{Quality and Performance versus Sample Rate}. Visual and performance results are compare to sample spacing $dt$.  Performance on the Orb model is <100 milliseconds when ray tracing high quality $2048^2$ images with $dt \leq 0.001$. Errors are observed primarily at grazing angles when $dt$ is large. We use $dt=0.002$ for visual results in Fig. \ref{fig:visual1}.}
  \label{fig:visual_sampling}
\end{figure}

 \section{Implementation}
Our algorithm is designed to integrate with standard hardware-based ray tracing engines that support bounding box, intersection and shading programs. We use Nvidia OptiX for efficient GPU-accelerated BVH trees \cite{Parker2010}\cite{Karras2013}. These are inserted into the BVH tree along with their AABBs (Fig. \ref{fig:overview}). 

\section{Results}
\begin{figure}[h!]
  \centering
  \includegraphics[width=0.75\textwidth]{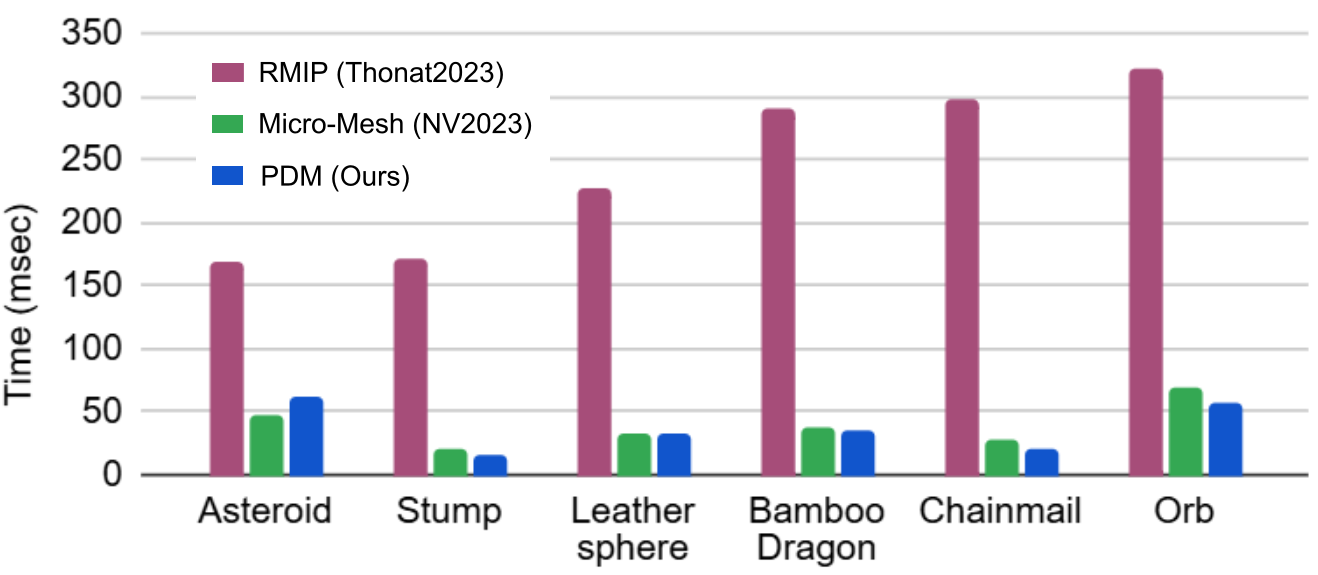}
  \caption{\textbf{Technique comparison by model}. RMIP (purple), Micro-Meshes (green) and our Projective Displacement Mapping (blue). Time per frame, 1 spp, 4 rays/pix. Lower is better.}
  \label{fig:graph_method}
\end{figure}

\begin{figure}[h!]
  \centering
  \includegraphics[width=0.75\textwidth]{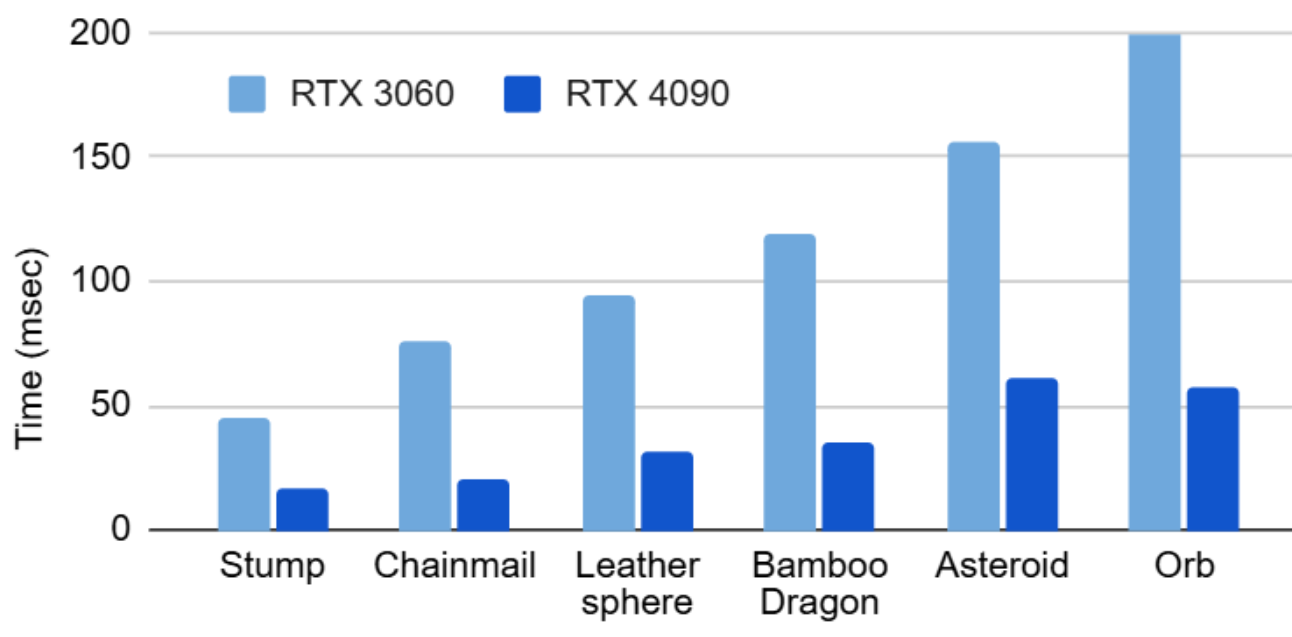}
  \caption{\textbf{Hardware comparison by model}. RTX 3060 (12.74 Gflops) versus RTX 4090 (82.6 Gflops).}
  \label{fig:graph_hw}
\end{figure}

Results are presented in several ways. Visual and performance comparisons on ray tracing were performed against both RMIP \cite{Thonat2023} and Micro-Meshes \cite{Nvidia2023} in Fig. \ref{fig:graph_method}. We also measure hardware deltas between a mid-range RTX 3060 and high-end RTX 4090 in Fig. \ref{fig:graph_hw}. Since our ray marching algorithm is limited by sample spacing we examine the visual quality and performance profile as $dt$ is decreased (Fig. \ref{fig:visual_sampling}). Memory is not compared here as there is no BLAS to evaluate; our method only requires the low-poly base mesh, 108 additional bytes per face for prism construction to store min/max extents, and the displacement map in GPU memory.

\subsection{Visual Results}
Visual comparisons are presented in Figures \ref{fig:visual_sampling} and \ref{fig:visual1}. Final images were rendered at $4096^2$ pixels with 64 samples consisting of 4 rays/pixel per sample: primary, path trace, reflection and shadow. For Micro-Meshes we modified the MicroMesh Toolkit source to output surface normal images. \cite{Thonat2023} provided surface normal images for RMIP comparisons.

A fur torus is rendered with a high frequency $4096^2$ displacement map. Thin features are well resolved by stochastic thin feature sampling, see Fig.  \ref{fig:visual2} and Section 5.3. There is no overhead for this technique since the same number of rays and samples are used. 

Although our technique is intended for authoring single objects, detailed scenes are possible with our method. Figure \ref{fig:visual_scene} contains nine displaced objects with different materials, rendered at 4096x1280 in 30 seconds with 462 ms/sample, 64 samples and 53.4 Mrays/sec while exhibiting four light sources, diffuse reflections, soft shadows and path tracing on a RTX 4090

\subsection{Performance}
Performance results are provided in Table 1. To quantify performance of algorithm stages we measured the average time for prism intersection $R_{isct}$, for just primary rays, and for ray tracing beauty images (4 rays/pix/sample). Our method ray traces on average between 20-60 milliseconds for $2048^2$ pixel images, with no editing overhead since we have no BLAS to rebuild.

Comparative performance to RMIP on our models was generously provided by \cite{Thonat2023}. Our method performed on average 40-60\% faster for primary rays and 2x-13x faster for beauty images in comparison to RMIP using identical hardware (RTX 4090). Although Micro-Meshes relies on tessellation, with goals and techniques different than ours, we still achieve faster ray tracing performance on most models tested. With the caveat that our method is not intended for massive scenes or compression of high polygon base meshes we achieve an overall performance between 80-400 Mrays/sec. 

Sample spacing is the limiting factor in our method. Visual quality and performance versus sample spacing are compared in Fig. \ref{fig:visual_sampling}.  Quality degrades primarily at grazing angles while we still achieve interactive rates under 100 ms on $2048^2$ images with $dt=0.001$ for all models.

\clearpage

\newcommand\highstrut{\leavevmode\raise\jot\copy\strutbox}
\newcommand\deepstrut{\leavevmode\lower\jot\copy\strutbox}

\begin{table}
  \label{tab:results}
  \setcellgapes{0pt}
  \begin{minipage}{\textwidth}
  \begin{center}
  \begin{tabular}{|l|r|r|r|r|r|r|r|r|r|r|}
  \hline 
  \highstrut & \highstrut & \multicolumn{2}{c|}{Prism Intersect} & \multicolumn{3}{c|}{Primary Rays (msec)} & \multicolumn{4}{c|}{Beauty Image (msec)}  \\
\hline
Model\highstrut & \# Tri. & $R_{isct}$ & $R_{isct}$ & Thonat & Ours & Improve & Micro- & Thonat & Ours & Improve \\
     &           & (msec) & \% & 2023 &  & \% & Mesh & 2023 & & \% \\   

\hline
\highstrut\textit{Asteroid} &\quad 480  & 11.7 & 19.4\% & 16.4 & 16.3 & 1\% & 48.3 & 145.6 &  60.8 &  139\%  \\
\textit{Chainmail} & 1423               & 11.8 & 57.3\% & 29.4 & 12.6 & 133\% & 28.9 & 289.0 &  20.7 & 1293\% \\
\textit{Leather Sphere} & 3968          & 13.2 & 41.4\% & 21.9 & 16.8 & 30\% & 33.7 & 230.6 &  31.9 &  622\% \\
\textit{Orb} & 3968                     & 13.2 & 23.3\% & 32.5 & 18.2 & 79\% & 68.2 & 329.1 &  56.8 &  479\%  \\
\textit{Tree Stump} & 7710              & 10.4 & 57.6\% & 18.9 & 13.3 & 42\% & 19.5 & 166.2 &  18.1 &  818\% \\
\textit{Bamboo Dragon} & 22308          & 13.0 & 36.6\% & 14.2 & 16.6 & -14\% & 37.7 & 282.9 &  35.7 &  692\% \\ 
\hline
  \end{tabular}
  \end{center}
  \end{minipage}    
  \vspace{1em}
  \caption{\textbf{Performance}. Comparisons of our method to RMIP \cite{Thonat2023} and Micro-Meshes \cite{Nvidia2023} for models of varying complexity on a GeForce RTX 4090. Displacement maps are 16-bit $4096^2$ pixels (Orb is $8192^2$). Beauty images consist of 64 samples at 4 rays/pixel: primary, path trace, reflection and shadow. All measurements are for ray tracing $2048^2$ output images at 1 sample/pixel and dt=0.002. }
\end{table}

\clearpage

\begin{figure*}
  \includegraphics[width=\textwidth]{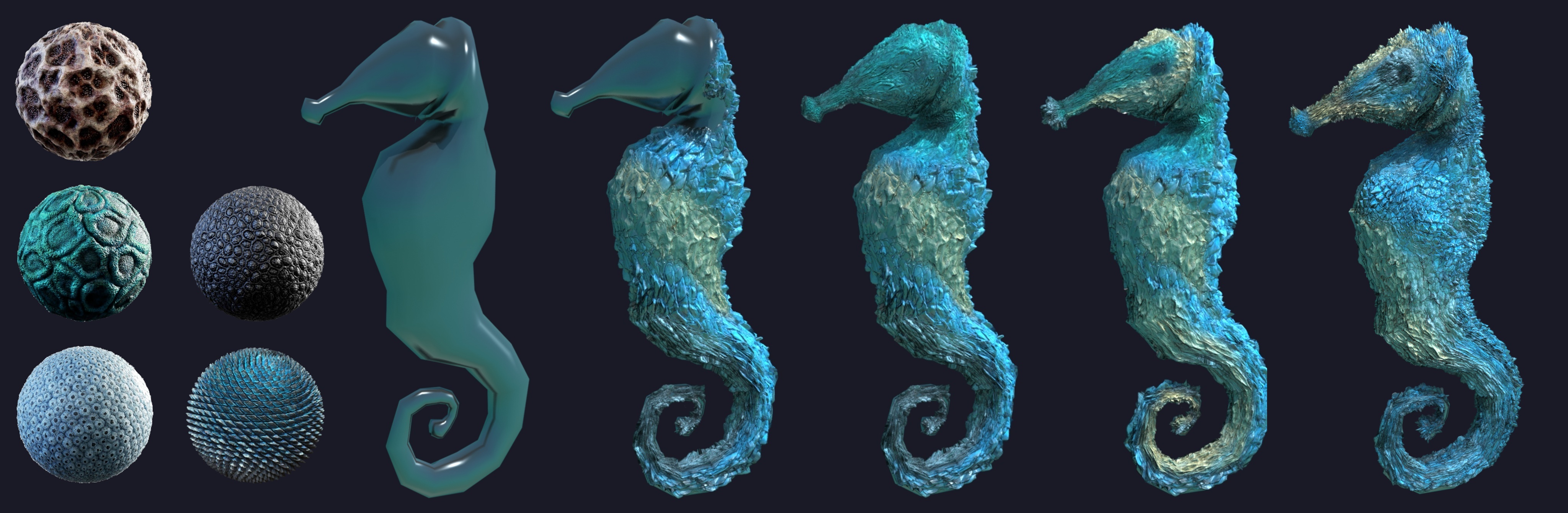}
  \caption{\textbf{Modeling example}. Detailed seahorse modeled interactively using our technique with displacement brushes while ray tracing.}
  \label{fig:seahorses}
\end{figure*} 

\subsection{Interactive Editing}
Displacement mapping is a tool for expressing detail for which artists desire interactive feedback. We developed a brush-like editor to demonstrate surface sculpting, see Figures \ref{fig:teaser} and \ref{fig:seahorses}. Surfaces can be sculpted with effects not previously possible such as editing with global illumination. A ray is traced at the mouse cursor on CPU with triangle-mesh intersection to find the barycentric and $u,v$ coordinates of a texel center. Displacement brushes are blending into the target displacement map with color and depth. During sculpting a region of $256^2$ pixels is ray traced around the mouse cursor at 28-34 msec (30 fps). Figure \ref{fig:seahorses} shows an example of authoring a $6400^2$ displacement map (see Supplemental video). We achieve interactive sculpting with no BLAS, regional updates and fast ray tracing. Upon mouse release or camera motion the full resolution image converges over multiple samples. These results allow artists to interactively edit fine details in 16-bit displacement textures with ray traced quality feedback, within the limits of the current displacement depth.

\begin{table}[h!]
  \label{tab:editing}
  \setcellgapes{0pt}
  \begin{minipage}{0.46\textwidth}
  \begin{center}
  \begin{tabular}{|l|r|r|r|r|r|}
  \hline 
  Method & \makecell[l]{Texture} & BLAS & \makecell[l]{Ray} & \makecell[l]{Total} & FPS \\
  & edits [1] & [2] & tracing [3] & (msec) & \\
\hline
\highstrut RMIP &\quad 3.2 & 3.4 & 329.1 & 335.7 & 2.9 \\
\highstrut PDM (full) &\quad 3.2 & 0 & 56.8 & 60.0 & 16.7 \\
\highstrut PDM ($256^2$) &\quad 3.2 & 0 & 11.0 & 14.2 & 70.4 \\
\hline
  \end{tabular}
  \end{center}
  \end{minipage}    
  \vspace{1em}
  \caption{\textbf{Editing performance}. Analysis of interative editing performance on RMIP and PDM (ours) for the Orb model. All times in milliseconds on RTX 4090. [1] Texture modifications to $8192^2$ displacement and color maps over a $256^2$ brush region, then transferred to GPU, [2] BLAS rebuild for a $1024^2$ rmip in \cite{Thonat2023}, none for ours, [3] Ray tracing beauty pass over modified pixels at 1 spp.}
\end{table}

Editing performance is analyzed in Table 2 for the $8192^2$ displaced Orb model. Each frame requires texture modifications to the displacement and color maps over the $256^2$ brush regions, updating BLAS data structures, and ray tracing at one sample per pixel. PDM (full) is our technique with full screen resampling, whereas PDM ($256^2$) is ours while updating a region of pixels around the cursor. As anticipated, texture and BLAS rebuild are minimal, whereas ray tracing is the dominant performance factor. 

\section{Conclusions}  
We present a new direct sampling technique for interactive ray tracing and editing of displacement mapped surfaces. Leveraging hardware BVHs for top-level acceleration (TLAS) we introduce a novel projective displacement algorithm for primitive intersection. Intentionally eliminating the bottom-level acceleration structure (BLAS) to reduce overhead and complexity, while improving the performance of direct sampling with new techniques, we are able to achieve interactive editing of displaced surfaces while ray tracing. Our CPU reference implementation is provided open source to facilitate future research.

Ray tracing by our method is faster than the existing BLAS based techniques RMIP and Micro-Meshes for low poly meshes with detailed displacement maps. Visually we improve on the appearance of displaced surfaces by eliminating faceting with a new smoothed displaced normal, by eliminating buckling with ray/bilinear patch prism interfaces, by stochastically sampling thin features, and by improving watertightness with prism extent analysis. We demonstrate the real-time sculpting of detailed displaced surfaces as seen through other reflective and refractive objects. Our method is efficient and simple, and integrates well with existing ray tracing engines.

Limitations of our technique are related to ray marching and sampling. This method is not intended for massive scenes, terrain, or compression of very high geometry static source meshes. We focus on the interactive design and sculpting of objects in look development work flows where memory resources may be limited. Beyond the input source mesh and displacement map our technique introduces minimal memory overhead for the prism BVH. No rebuild is needed per frame if the displacement edits are less than a fixed maximum offset distance, whereas larger changes to the base mesh would require BVH reconstruction.

In the future we hope to examine the trade-offs between direct sampling and the rebuilding of BLAS acceleration structures while attempting to maintaining interactivity. We would also like to explore $C^1$ intermediate surfaces to further study aspects of continuity. The geometric detail and natural authoring of displacement maps present compelling reasons for further exploration. We hope this work expands future opportunities for complex geometric modeling and interactive ray tracing.

\section{Acknowledgements}
Special thanks to Theo Thonat and Adobe Research for performance, visual results and discussions on RMIP, and to Pyralel Knowles and Nvidia for discussions on Micro-Meshes. Tree Stump model by Quixel Megascans. Displacement textures for orb, fish scales and creature skins by CGAxis.com. All other textures are public domain, CC0, or by the author.

\clearpage

%% Appendicies
\begin{figure*}
  \centering
  \includegraphics[width=0.85\textwidth]{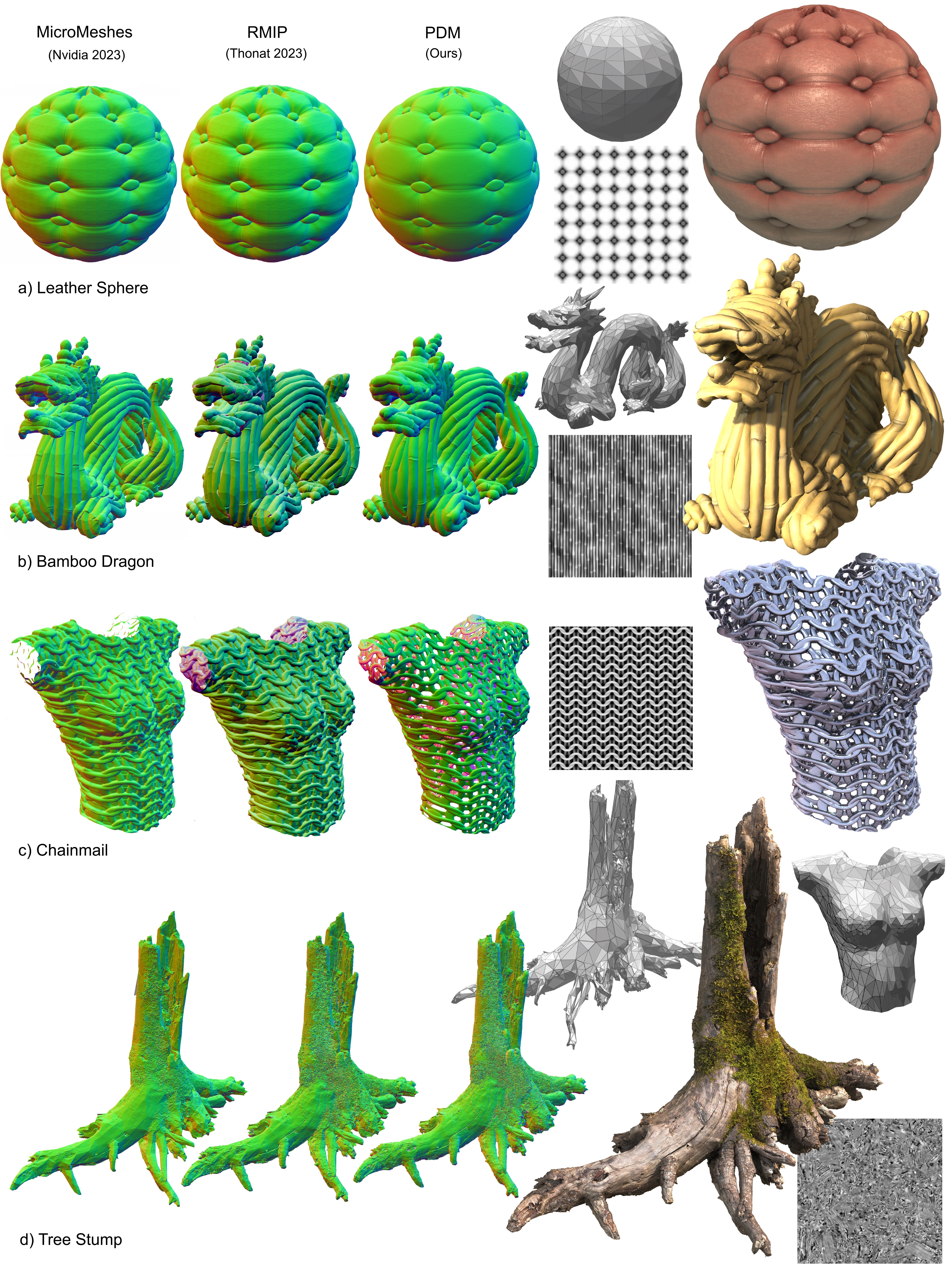}
  \caption{\textbf{Ray tracing results}. Displaced surface normals are compared between Micro-Meshes \cite{Nvidia2023}, RMIP \cite{Thonat2023} and our Projective Displacement method with $2048^2$ pixel at 64 spp (left). The base mesh, displacement map, and beauty images are shown for each model (right). The smooth surfaces in our leather sphere, dragon and tree stump are due to surface normal correction (Sec. 5.1), and the chainmail correctly renders holes as we use the alpha channel of the color texture to determine opacity.}
  \label{fig:visual1}
\end{figure*}

\begin{figure*}
  \centering
 \includegraphics[width=0.85\textwidth]{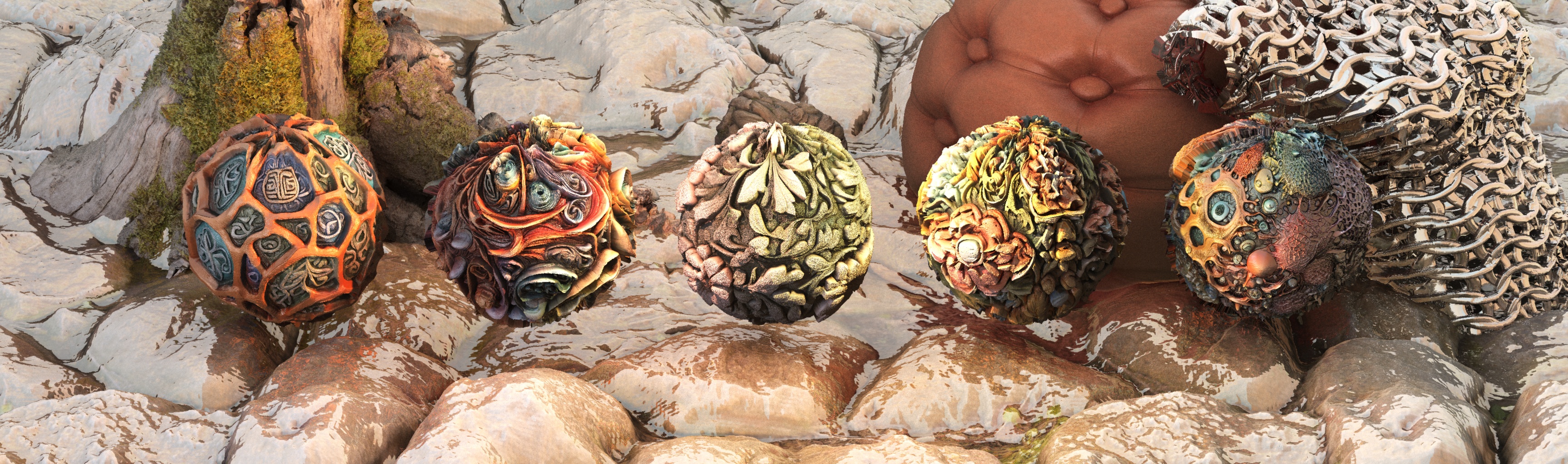}
 \caption{\textbf{Detailed scene}. A detailed scene ray traced with only 15k triangles and 1.8 MB GPU memory in 30 seconds at 4096x1280. All objects use our technique except for the ground surface.}
  \label{fig:visual_scene}
\end{figure*}

\begin{figure*}
  \centering
  \includegraphics[width=0.85\textwidth]{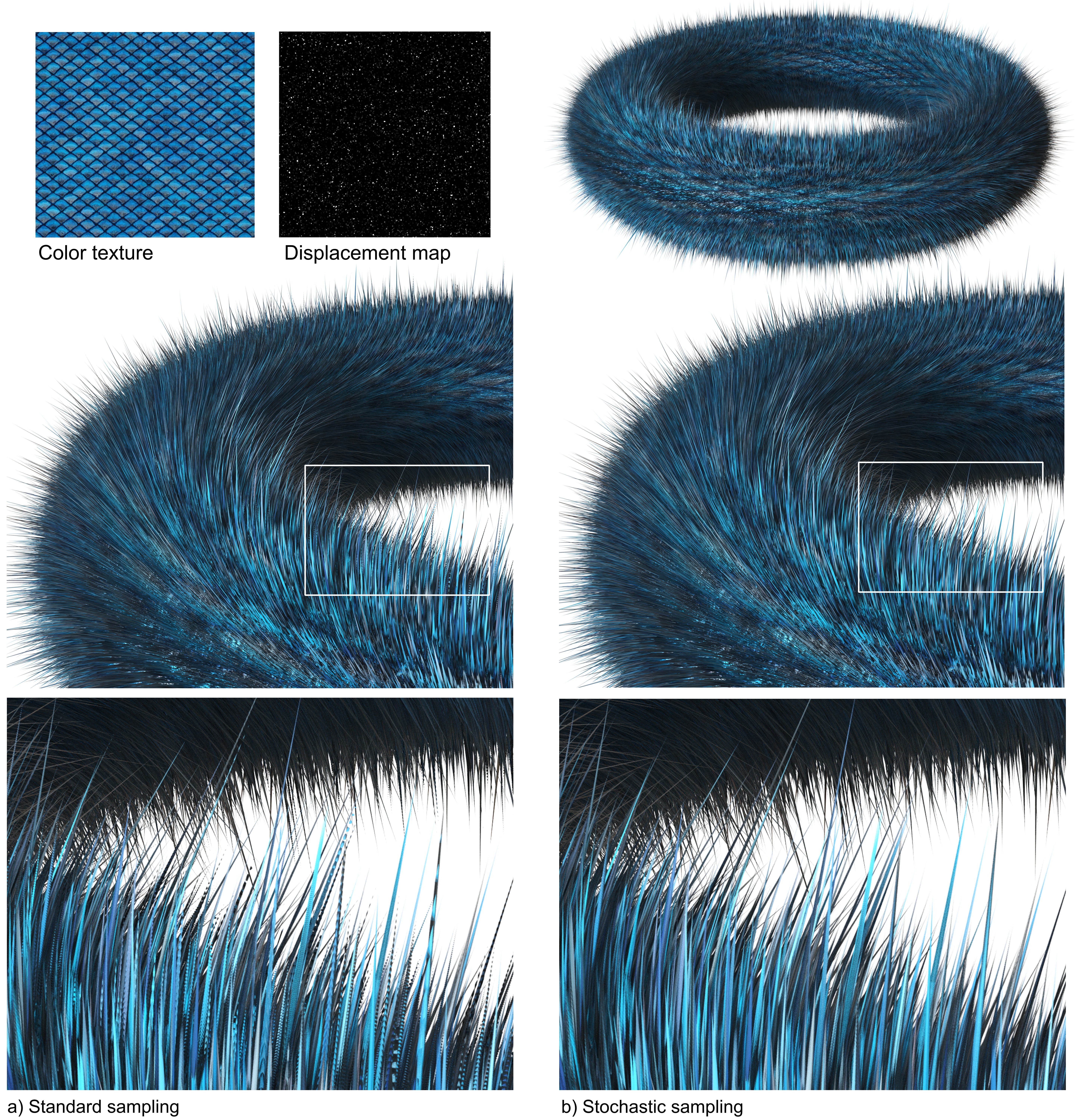}
  \caption{\textbf{Thin feature sampling}. Thin features are ray traced with stochastic sampling, see Section 5.3. The fur torus is rendered at $4096^2$, 64 spp, and $dt=0.001$. a) Uniform sampling causes thin features to be periodically missed, whereas b) Stochastic sampling uses distributed first samples along $t$ after prism intersection to integrate over these features. This method incurs no overhead since the same number of rays and samples are used.}
  \label{fig:visual2}
\end{figure*}

\appendix
\label{apxA}

\clearpage

\Large Appendices
\normalsize

\setcounter{equation}{0}

\section{Barycentric Points Are Colinear with Interpolated Normals}

\begin{proof}
Show that the barycentric coordinates \((u,v,1-u-v)\) for points in an offset triangle $\mathbf{T}(u,v,w)$ reside on the interpolated surface normal $\mathbf{N}'$ of a displaced surface $S$.

Given a set of extruded, offset triangles defined along interpolated normals by: 
\begin{equation}
\begin{aligned}
    \mathbf{N}'(u,v) = \left| u \mathbf{n}_0 + v \mathbf{n}_1 + (1-u-v) \mathbf{n}_2 \right| \\
    \mathbf{T}(u,v,w) = \mathbf{P}(u,v) + w \mathbf{N}'(u,v)
\end{aligned}
\end{equation}

And given the displaced surface defined by:
\begin{equation}
    \mathbf{S}(u,v,w) = \mathbf{P}(u,v) + D(u,v) \mathbf{N}'(u,v)
\end{equation}

We intend to prove that for all values of \( w \) the points \(\mathbf{T}(u,v,w)\) reside along the normal \(\mathbf{N}'(u,v)\), which only requires demonstrating the points are co-linear and parallel to \(\mathbf{N}(u,v)\). Co-linearity is proven by showing that \(\mathbf{T} - \mathbf{P}\) is a scalar multiple of \(\mathbf{N}\). From the definition of offset triangles above:
\begin{equation}
    \mathbf{T}(u,v,w) - \mathbf{P}(u,v) = w \mathbf{N}'(u,v)
\end{equation}

The scalar multiple \(w\) \textit{is} the multiple of \(\mathbf{N}\) which gives \(\mathbf{T} - \mathbf{P}\), therefore, \(\mathbf{T}(u,v,w)\) resides along \(\mathbf{N}\) for all points on a given \(u,v\). Since $\mathbf{S}$ is defined with the same base surface $\mathbf{P}$ and normals $\mathbf{N}$, points in offset triangles $\mathbf{T}$ are on the same normals as the displaced surface $\mathbf{S}$.
\end{proof}

\section{Displaced and Corrected Surface Normals}

\begin{proof}

Construct a smoothed displaced surface normal $\mathbf{N}_s'$ that contains the interpolated (smoothed) base triangle normal $\mathbf{N}'$ for base triangles $\mathbf{P}$, and scalar displacement function $D$.

We show first that the evaluated normal $\mathbf{N}_s$ of a displaced surface $\mathbf{S}$, sampled at a point \((u,v)\) on a triangle using finite differences, contains a term for the flat geometric normal $\mathbf{N}_g$ of the base triangles even though the surface is constructed using the interpolated normal $\mathbf{N}'$.

Given the displaced surface:
\begin{equation}
    \mathbf{S}(u,v,w) = \mathbf{P}(u,v) + D(u,v) \mathbf{N}'(u,v)
\end{equation}

The surface normal is given by the cross product of the tangent and bi-tangent:
\begin{equation}
    \mathbf{N}_s = \frac{d\mathbf{S}}{du} \times \frac{d\mathbf{S}}{dv}
\end{equation}

Therefore, taking the partial derivatives of equation (4):
\begin{equation}
    \frac{d\mathbf{S}}{du} = \frac{d\mathbf{P}}{du} + \frac{dD}{du} \mathbf{N}'
\end{equation}
\begin{equation}
    \frac{d\mathbf{S}}{dv} = \frac{d\mathbf{P}}{dv} + \frac{dD}{dv} \mathbf{N}'
\end{equation}

Rewriting equation (5) in terms of these:
\begin{equation}
    \mathbf{N}_s = \left( \frac{d\mathbf{P}}{du} + \frac{dD}{du} \mathbf{N}' \right) \times \left( \frac{d\mathbf{P}}{dv} + \frac{dD}{dv} \mathbf{N}' \right)
\end{equation}

Using the distributive property of the cross product, ie. $(\mathbf{A}+\mathbf{B}) \times (\mathbf{C}+\mathbf{D}) = (\mathbf{A} \times \mathbf{C}) +(\mathbf{A} \times \mathbf{D}) +(\mathbf{B} \times \mathbf{C}) +(\mathbf{B} \times \mathbf{D})$:
\begin{equation}
    \mathbf{N}_s = \left( \frac{d\mathbf{P}}{du} \times \frac{d\mathbf{P}}{dv} \right) + \left( \frac{d\mathbf{P}}{du} \times \frac{dD}{dv} \mathbf{N}' \right) +
    \left( \frac{dD}{du} \mathbf{N}' \times \frac{d\mathbf{P}}{dv} \right) + 
    \left( \frac{dD}{du} \mathbf{N}' \times \frac{dD}{dv} \mathbf{N}' \right)
\end{equation}

The first term is just the geometric (flat) normal of the base triangle:
\begin{equation}
    \left( \frac{d\mathbf{P}}{du} \times \frac{d\mathbf{P}}{dv} \right) = \mathbf{N}_g
\end{equation}

The next two terms in (9) can be rewritten using the property of cross product scalar factoring, i.e., \(s_1 \mathbf{A} \times s_2 \mathbf{B} = s_1 s_2 (\mathbf{A} \times \mathbf{B})\):
\begin{equation}
    \left( \frac{d\mathbf{P}}{du} \times \frac{dD}{dv} \mathbf{N}' \right) = \frac{dD}{dv} \left( \frac{d\mathbf{P}}{du} \times \mathbf{N}' \right)
\end{equation}
\begin{equation}
    \left( \frac{dD}{du} \mathbf{N}' \times \frac{d\mathbf{P}}{dv} \right) = \frac{dD}{du} \left( \mathbf{N}' \times \frac{d\mathbf{P}}{dv} \right)
\end{equation}

Where the first part, $( \frac{dD}{du}, \frac{dD}{dv} )$, is the gradient $\nabla D(u,v)$ of the displacement texture, and the second part is the directional derivative of the surface normal $\mathbf{N}'$ across the base surface $\mathbf{P}$.

The last term in equation (9) can also be reduced by cross product scalar factoring:
\begin{equation}
    \left( \frac{dD}{du} \mathbf{N}' \times \frac{dD}{dv} \mathbf{N}' \right) = \frac{dD}{du} \frac{dD}{dv} \left( \mathbf{N}' \times \mathbf{N}' \right)
\end{equation}

Since the cross product of a vector with itself is zero, we have:
\begin{equation}
    \left( \mathbf{N}' \times \mathbf{N}' \right) = 0
\end{equation}

Rewriting equation (9) using equations (10), (11), (12), and (14), we have:
\begin{equation}
    \mathbf{N}_s = \mathbf{N}_g + \frac{dD}{dv} \left( \frac{d\mathbf{P}}{du} \times \mathbf{N}' \right) + \frac{dD}{du} \left(\mathbf{N}' \times \frac{d\mathbf{P}}{dv} \right)
\end{equation}
    
Or more compactly:
\begin{equation}
    \mathbf{N}_s = \mathbf{N}_g + \nabla D(u,v) \otimes \nabla_{\mathbf{P}} \mathbf{N}'(u,v)
\end{equation}

We note that \(\mathbf{N}_g\) is the geometric, flat normal of the base triangle, while the other term is directional derivative of the interpolated normal $\mathbf{N}'$ across the surface $\mathbf{P}$ scaled by the gradient of the displacement map $D$. This demonstrates that the evaluated surface normal \(\mathbf{N}_s\) of a displaced surface over a set of base triangles contains the flat geometric normal \(\mathbf{N}_g\), despite being defined over the interpolated normal $\mathbf{N}'$.

We wish to construct a smoothly interpolated displaced surface normal $\mathbf{N}'_s$. Construction follows the idea of Phong shading, which replaces the geometric normal \(\mathbf{N}_g\) with the smooth interpolated normal \(\mathbf{N}'\) when rendering non-displaced triangles. For displaced surfaces we can replace it as:

\begin{equation}
    \mathbf{N_s}' = \mathbf{N}_s - \mathbf{N}_g + \mathbf{N}' 
\end{equation}
    
Substituting $\mathbf{N}_s$ from (16) into equation (17).
\begin{align}
    \mathbf{N_s}' &= \left[ \mathbf{N}_g + \nabla D(u,v) \otimes \nabla_{\mathbf{P}} \mathbf{N}'(u,v) \right] - \mathbf{N}_g + \mathbf{N}' \\
    \mathbf{N_s}' &= \mathbf{N}' + \nabla D(u,v) \otimes \nabla_{\mathbf{P}} \mathbf{N}'(u,v)
\end{align}

Equations (17) may be used if $\mathbf{N}_s$ is already given through numerical evaluation, or equation (19) when the base triangle $\mathbf{P}$, the displaced texture $\mathbf{D}$, and the interpolated normal $\mathbf{N}'$ are available. This last equation (19) is identical to equation (16) except that $\mathbf{N}_g$ has been replaced by $\mathbf{N}'$.

\end{proof}

\section{Barycentric Coordinates in a Triangle}
The scanning step of our Projective Displacement Mapping computes triangle barycentric coordinates of a point $s$ inside a scanning triangle $c_i$. The function \textit{triangleBarycentric} is based on \cite{Ericson2004} with minor performance improvements:

\label{alg:two}
\begin{verbatim}
float3 triangleBarycentric (float3 s,
  float3 c0, float3 c1, float3 c2)
{
    float3 x0, x1, x2, b;
    float d00, d01, d11, d20, d21, invDenom;
    x0 = c1-c0; x1 = c2-c0; x2 = s - c0;
    d00 = dot(x0, x0);
    d01 = dot(x0, x1);
    d11 = dot(x1, x1);
    d20 = dot(x2, x0);
    d21 = dot(x2, x1);
    invDenom = 1.0 / (d00 * d11 - d01 * d01);
    b.y = (d11 * d20 - d01 * d21) * invDenom;
    b.z = (d00 * d21 - d01 * d20) * invDenom;
    b.x = 1.0-b.y-b.z;
    return b;
}
\end{verbatim}

\clearpage

\bibliography{ms}  % Uncomment to use external .bib file

\end{document}